\documentclass[preprint,aps,nofootinbib,showpacs,amsfonts,epsf]{revtex4-2}

\input{epsf.tex}

\newcommand{\E}{{\cal{E}}}
\newcommand{\s}{\sigma}

\renewcommand{\a}{\alpha}

\newcommand{\be}{\begin{equation}}
\newcommand{\ee}{\end{equation}}
\newcommand{\bea}{\begin{eqnarray}}
\newcommand{\eea}{\end{eqnarray}}
\newcommand{\ba}{\begin{array}}
\newcommand{\ea}{\end{array}}
\def\J#1#2#3#4{{#1} {\bf #2}, #3 (#4)}
\def\PRD{Phys. Rev. D}
\def\PR{Phys. Rev.}
\def\PRL{Phys. Rev. Lett.}
\def\PTP{Prog. Theor. Phys.}

\def\JMP{J. Math. Phys.}

\def\CMP{Commun. Math. Phys.}
\def\MZ{Math. Z.}

\def\CQG{Class. Quantum Grav.}

\def\PLA{Phys. Lett. A}

\def\NPB{Nucl. Phys. B}
\def\MZ{Math. Zeits.}
\def\NC{Nuovo Chim.}

\def\JHEP{J. High Energy Phys.}

\begin{document}
\draft
\title{The temperature and free energy of multi-black hole systems}

\author{C. J. Ram\'irez-Valdez, H. Garc\'ia-Compe\'an, and V.~S.~Manko}
\address{Departamento de F\'\i sica, Centro de Investigaci\'on y
de Estudios Avanzados del IPN, A.P. 14-740, 07000 Ciudad de
M\'exico, Mexico}

\begin{abstract}
In the present paper, we compute the Euclidean action of a generic
system consisting of $N$ arbitrary Kerr-Newman black holes located
on the symmetry axis and separated from each other by massless
struts. This allows us to introduce the {\it Hawking average
temperature} (HAT) $\hat T$ of the multi-black hole system via the
condition of vanishing the entire set of terms involving the
singular horizons due to periodic time, and the resulting formula
for this temperature contains solely the surface gravities
$\kappa_i$ and horizon areas $A^H_i$ of the black hole constituents.
We also show that the corresponding expression for the {\it free
energy} of the system defined by $\hat T$ is consistent with the
first law of thermodynamics and Smarr mass relations.
\end{abstract}

\pacs{04.20.Jb, 04.70.Bw, 97.60.Lf}

\maketitle

\newpage

\section{Introduction}

The Euclidean action approach to black hole physics was introduced
by Gibbons and Hawking in their seminal paper \cite{GHa} where they
confirmed in an elegant new way the previously found analytic
relations of the Hawking temperature to the surface gravity
\cite{Haw}, and of the black hole's entropy to the area of the event
horizon \cite{Bek}. Though this approach was originally applied to
the study of the thermodynamic behavior of a single black hole, the
subsequent interest in the binary black-hole configurations,
stimulated by the development of modern solution generating
techniques and construction of various multi-black hole metrics
\cite{KNe,MMR}, later led to the appearance of a number of papers in
which the Euclidean formalism was employed for the analysis of some
specific double-black-hole systems with {\it equal} surface
gravities of the constituents. Thus, in the paper \cite{CPe}, Costa
and Perry evaluated the free energy of two equal Schwarzschild black
holes and, compared to the paper \cite{GHa}, the action integral of
the binary system had an additional term arising from the massless
strut \cite{Isr} separating the constituents. The Euclidean action
of a pair of identical (up to a sign of charges)
Reissner-Nordstr\"om nonextreme black holes was calculated by
Emparan and Teo \cite{ETe}, while the thermodynamical properties of
equal Kerr black holes, with a brief analysis of the action
integrals, were studied numerically by Herdeiro {\it et al.}
\cite{HRR}. In our recent paper \cite{GMR} we have computed the
Euclidean action for two different binary systems of equal
counterrotating Kerr-Newman (KN) black holes and obtained the
corresponding expressions for the free energy of each system. Note
that in the above papers \cite{CPe,ETe,HRR,GMR} all the binary
systems were in thermal equilibrium because the black holes in them
had the same surface gravities and hence the same Hawking
temperature $T_H$.

In the paper \cite{GMR} we have observed that it is quite simple to
guess the explicit expression of the free energy, which is closely
related to the Euclidean action, in the general case of the
double-Schwarzschild solution \cite{BWe} from the respective formula
obtained in the special case of equal Schwarzschild black holes.
This naturally raises the question of whether the Euclidean action
method can be extended to the systems of arbitrary black holes with
{\it nonequal} surface gravities, similar to thermodynamics which
goes far beyond the black hole configurations in thermal equilibrium
\cite{CGLP}. The reason of why such an extension of the method has
not yet been attempted up to now is in fact known and looks quite
meaningful at first sight -- a Wick rotation of the multi-black hole
spacetime introduces conical singularities at the horizons due to
the periodic Euclidean time $\tau$, and whereas one conical
singularity, say, on the first horizon can be eliminated by choosing
appropriately the period of $\tau$, the other horizons will remain
singular unless they have the same surface gravities as the first
horizon. However, as will be shown in the present paper, the way out
of this seemingly unresolvable situation consists in calculating the
entire Euclidean action, including the terms arising from all the
singular horizons; then the condition of vanishing the combined
contribution of the latter `singular horizon' terms will fix the
value of the period of $\tau$, the inverse of which will give the
{\it Hawking average temperature} (HAT) of the system. It is
remarkable that the temperature introduced in this way turns out to
be determined by a very concise formula involving exclusively the
horizons' areas and surface gravities; moreover, the corresponding
{\it free energy} of the system becomes consistent with the first
law of thermodynamics and the well-known Smarr relations \cite{Sma}.

This paper is organized as follows. In the next section we will
reexamine the derivation of the Euclidean action in the case of a
single KN black hole \cite{NCC}, evaluating explicitly the
contribution of the conical singularity due to Euclidean time. In
Sec.~III we calculate the Euclidean action for a system of $N$
collinear arbitrary KN black holes described by the extended
$N$-soliton electrovac solution \cite{RMM}. Here two different types
of the conical singularity contributions will appear - the one
coming from the singular horizons and the other arising from the
struts. The Hawking average temperature $\hat T$ is introduced in
Sec.~IV where we find its form in terms of the surface gravities and
horizons' areas of black holes; we also give the expression of the
free energy $W$ defined by $\hat T$ and show its consistency with
both the first law of thermodynamics and Smarr's mass formula.
Sec.~V contains concluding remarks.

Throughout the paper, units are used in which $c=G=\hbar=k_B=1$.

\section{Euclidean action of a single Kerr-Newman black hole}

The Euclidean action we are interested in can be represented in the
form \cite{GHa}
\be\label{action}
I_{\rm E}=I_{\rm EH}+I_{\rm GH}+I_{\rm em}, \ee
where the first term $I_{\rm EH}$ is the well-known Einstein-Hilbert
action in Euclidean signature
\be\label{ieh} I_{\rm
EH}=-\frac{1}{16\pi}\int_{\mathcal{M}}R\sqrt{g}, \ee
the second term $I_{\rm GH}$ is the Gibbons-Hawking boundary term
\be\label{igh}
    I_{\rm GH}=-\frac{1}{8\pi}\int_{\partial\mathcal{M}}[K]\sqrt{h},
\ee
with $[K]:=K-K_0$, and the last term $I_{\rm em}$ is the standard
electromagnetic action in the absence of currents
\be\label{iem}
    I_{\rm em}=\frac{1}{16\pi}\int_{\mathcal{M}} F\wedge\star F.
\ee

The {\it free energy} of the system is related to $I_{\rm E}$ by the
simple formula
\be\label{fe}
     W=I_{\rm E}/\beta, \quad \beta=1/T,
\ee
where $\beta$ is the period of the Euclidean time $\tau=it$, and $T$
denotes the temperature.

In the presence of a conical singularity, the action integral
$I_{\rm EH}$ can be evaluated with the aid of the formula
\cite{CPe,Reg,FSo}
\be\label{cs}
    \frac{1}{2}\int_\mathcal{M} R\sqrt{g}=\mathit{Area}\cdot\delta,
\ee
where $Area$ is the area of the surface spanned by the conical
singularity, and $\delta$ is the angle deficit.

As for the Gibbons-Hawking boundary term $I_{\rm GH}$ that must be
computed at spatial infinity, it can be seen that it consists of two
parts. The first one, namely,
\be\label{K}
    \int_{\partial\mathcal{M}} K \sqrt{h},
\ee
which involves the extrinsic curvature $K$ of $\partial\mathcal{M}$,
is intrinsically divergent and requires a counter-term for
regularization. The latter term is given by the second part of
$I_{\rm GH}$, namely,
\be\label{K0}
    \int_{\partial\mathcal{M}} K_0 \sqrt{h},
\ee
where $K_0$ is the extrinsic curvature of the same surface
$\partial\mathcal{M}$ in flat space. One possible way to compute
(\ref{K}) is by using formula \cite{GHa,CPW}
\be\label{Kn}
  \int_{\partial\mathcal{M}} K\sqrt{h}=
  n\left({\rm Vol}\,\partial\mathcal{M}\right),
\ee
where $n$ is a unit normal vector field over $\partial\mathcal{M}$,
while ${\rm Vol}\,\partial\mathcal{M}$ stands for the volume of
$\partial\mathcal{M}$.

We find it instructive, before treating the general case of $N$
arbitrary KN black holes, first to illustrate the use of the above
formulas by the example of a single KN black hole. A Wick rotated KN
metric, written in Boyer-Lindquist coordinates, has the form
\bea\label{KN}
    ds^2&=&\left(1-\frac{2mr-q^2}{\Sigma}\right)d\tau^2
    +\frac{\Sigma}{\Delta}dr^2+\Sigma d\theta^2+\frac{C}
    {\Sigma}\sin^2{\theta}d\varphi^2+\frac{2ia(2mr-q^2)}
    {\Sigma}\sin^2{\theta}d\tau d\varphi, \nonumber\\
\Sigma&=&r^2+a^2\cos^2{\theta}, \quad
    \Delta=r^2-2mr+a^2+q^2, \nonumber\\
    C&=&\left(r^2+a^2\right)^2-a^2\Delta\sin^2{\theta},
\eea
the parameters $m$, $a$ and $q$ representing, respectively, the
mass, angular momentum per unit mass and electric charge of the KN
black hole.

We start the calculation of the Euclidean action of the KN solution
with the evaluation of the term $I_{\rm EH}$ via formula (\ref{cs}).
The area spanned by the conical singularity at the horizon is
precisely the horizon's area $A^H$, while for the conical deficit
$\delta^\tau$ in terms of the surface gravity $\kappa$ we obtain
\be\label{del}
    \delta^\tau=2\pi-\int_0^\beta \kappa d\tau
    =2\pi\left(1-\frac{\kappa}{2\pi}\beta\right),
\ee
whence it follows that
\be\label{IehKN}
     I_{\rm EH}=-\frac{1}{8\pi}A^H\delta^\tau,
\ee
the concrete well-known values of $A^H$ and $\kappa$ of the KN black
hole being
\be\label{Ak} A^H=4\pi(r_+^2+a^2), \quad
\kappa=\frac{r_+-r_-}{2(r_+^2+a^2)}, \quad
r_\pm=m\pm\sqrt{m^2-a^2-q^2}. \ee

Taking into account the positivity of $A^H$, it is clear that the
only possibility to get rid of the conical singularity contribution
and achieve $I_{\rm EH}=0$ would be demanding $\delta^\tau=0$, which
implies
\be\label{Tk}
    \beta=\frac{2\pi}{\kappa} \quad \Rightarrow \quad
    T=\frac{\kappa}{2\pi},
\end{equation}
so that the removal of the conical singularity due Euclidean time
automatically fixes the temperature $T$ of the KN black hole in the
Hawking form (\ref{Tk}).

Turning now to the evaluation of the Gibbons-Hawking term $I_{\rm
GH}$, we must bear in mind that the boundary lies at spatial
infinity where each of the integrals (\ref{K}) and (\ref{K0})
diverges. Therefore, on the one hand, we must first compute the
integrals for some finite $r$ and then take the limit $r\to\infty$
in the final expression for $I_{\rm GH}$; on the other hand, we can
use the large-$r$ approximation to slightly simplify the
computational process.

Let $\mathcal{N}$ be a hypersurface defined by $r={\rm const}$. Then
the determinant $h$ of the induced metric over $\mathcal{N}$ for
large $r$ takes the form
\be\label{h}
   h\simeq \left(1-\frac{2m}{r}\right)r^4\sin^2{\theta},
\ee
and we get for the volume of $\mathcal{N}$:
\be\label{VN}
   {\rm Vol}\,\mathcal{N}=\int^\pi_0\int^{2\pi}_0\int^{\beta}_0
   \sqrt{h}\ d\theta d\varphi d\tau
    \simeq 4\pi\beta r^2\left(1-\frac{m}{r}\right).
\ee

If we consider now a normalized vector field $n$ over $\mathcal{N}$
defined as $g_{rr}^{-1/2}\partial_r$, then in our approximation
\be\label{n}
    n\simeq\left(1-\frac{m}{r}\right)\partial_r,
\ee
so that
\be\label{nVN}
    n\left({\rm Vol}\;\mathcal{N}\right)
    \simeq 4\pi\beta(2r-3m),
\ee
and
\be\label{Kh}
    -\frac{1}{8\pi}\int_\mathcal{N} K\sqrt{h}
    \simeq -\frac{1}{2}\beta(2r-3m).
\ee

On the other hand, the extrinsic curvature of $\mathcal{N}$ in flat
space is just $K_0=2/r$, so we get
\be\label{K0h}
    -\frac{1}{8\pi}\int_\mathcal{N} K_0\sqrt{h}
    \simeq -\frac{1}{2}\beta\left(1-\frac{m}{r}\right)2r,
\ee
and the combination of (\ref{Kh}) and (\ref{K0h}) yields
\be\label{KK0}
    -\frac{1}{8\pi}\int_\mathcal{N} \left[K\right]\sqrt{h}
    =\frac{1}{2}\beta m+O\left(r^{-1}\right).
\ee

Taking the limit $r\to\infty$ in (\ref{KK0}), we finally arrive at
the Gibbons-Hawking term $I_{\rm GH}$ of the KN black hole:
\be\label{GHKN}
    I_{\rm GH}=\frac{1}{2}\beta m.
\ee

To find the remaining part of the Euclidean action related to the
electromagnetic field, we shall employ the approach earlier used in
the papers \cite{ETe,CPW} to avoid manipulations with the electric
potential on the horizon made in the original paper \cite{GHa}. In
fact, this calculational procedure is similar to the one just
employed for the evaluation of the Gibbons-Hawking term $I_{\rm GH}$
-- we shall perform the computations for some generic finite value
of $r$, but at the end, instead of the limit $r\to\infty$, we shall
take the limit $r\to r_+$ corresponding to the horizon of the KN
black hole. This in particular excludes the use of any approximation
tricks during the calculations.

In the Euclidean signature, the electromagnetic potential of the KN
solution has the form
\be\label{Aem}
    A=A_{\tau}(r,\theta)d\tau+A_{\varphi}(r,\theta)d\varphi
    =\frac{qr}{\Sigma}(id\tau+a\sin^2{\theta}d\varphi),
\ee
and these components of $A$ must be substituted into the integral
(\ref{iem}) which can be readily rewritten as a boundary term
\be\label{boun}
     \int_\mathcal{M} F^{\mu \nu}F_{\mu\nu}\sqrt{g}\;d^4x
     =2\int_{\partial\mathcal{M}} F^{\nu\mu}A_\mu n_\nu
     \sqrt{h}\;d^3x.
\ee

As before, the integral on the right-hand side of (\ref{boun}) will
be computed over a hypersurface $\mathcal{N}$ defined by $r={\rm
const}$, with the same normal unit vector $n$, and the integration
is straightforward:
\bea\label{intem}
    \int_\mathcal{N} F^{\nu\mu}A_\mu n_\nu \sqrt{h}d^3x
    &=&2\pi\beta \int^\pi_0 \frac{d\theta}{\sin{\theta}}
    \left[\left(g_{\varphi\tau}A_\varphi-g_{\varphi\varphi}A_\tau\right)
    \partial_r  A_\tau+\left(g_{\tau\varphi}A_\tau-g_{\tau\tau}A_\varphi\right)
    \partial_r  A_\varphi\right] \nonumber\\
    &=&2\pi\beta q^2r\int^\pi_0\frac{\Sigma-2r^2}{\Sigma^2}\sin{\theta}d\theta
    \nonumber\\
    &=&2\pi\beta q^2r\partial_r\int^\pi_0\frac{r}{\Sigma}\sin{\theta}d\theta
    \nonumber\\
    &=&4\pi\beta q^2r\partial_r\arctan(a/r) \nonumber\\
    &=&-4\pi\beta \frac{q^2r}{r^2+a^2}\label{r-potential}.
\eea

Taking now the limit $r\to r_+$, we immediately obtain
\be\label{lim}
 \lim_{r\to r_+} \int_\mathcal{N} F^{\nu\mu}A_\mu n_\nu \sqrt{h}\,d^3x
    =-4\pi\beta \frac{q^2r_+}{r_+^2+a^2}
    =-4\pi\beta q\Phi,
\ee
where the value of the electric potential $\Phi$ on the horizon is
determined by the formula
\be\label{Phi} \Phi=\frac{qr_+}{r_+^2+a^2}. \ee

Therefore, for the electromagnetic action integral $I_{\rm em}$ we
finally get
\be\label{iemKN}
    I_{\rm em}=-\frac{1}{2}\beta q\Phi,
\ee
so that the entire Euclidean action $I_{\rm E}$ takes the form
\be\label{IE} I_{\rm E}=\frac{1}{2}\beta(m-q\Phi)
-\frac{1}{8\pi}A^H\delta^\tau, \ee
with $A^H$ and $\delta^\tau$ defined by (\ref{Ak}).

As it has already been mentioned, the choice of $\beta$ in the form
(\ref{Tk}) causes vanishing of the last term in (\ref{IE}), which is
required for the regularity of the horizon. Thus, when
$\beta=1/T=2\pi/\kappa$, the free energy of the KN black hole, as it
follows from (\ref{fe}), is given by
\be\label{wKN} W=\frac{1}{2}(m-q\Phi). \ee

We now turn to the discussion of the general case of $N$ black
holes.

\section{Euclidean action of $N$ Kerr-Newman black holes}

The configuration of $N$ collinear arbitrary KN black holes is
described by a subfamily of the extended $N$-soliton electrovac
solution \cite{RMM} constructed with the aid of Sibgatullin's
integral method \cite{Sib}. It is curious that a year after the
publication of the paper \cite{RMM} with a concise explicit form of
all the metrical fields, an article appeared in a mathematical
journal \cite{Wei} in which solely the proof of the existence and
uniqueness of the Ruiz {\it et al.} $N$-soliton solution was
attempted. We recall that the Ernst potentials \cite{Ern} of the
solution \cite{RMM} are defined by the expressions
\bea\label{EF} &&{\cal E}=E_+/E_-, \qquad \Phi=F/E_-, \nonumber\\
&&E_\pm=\left|\begin{array}{cccc} 1 & 1 & \ldots & 1 \\ \pm 1 &
{\displaystyle \frac{r_1}{\a_1-\beta_1}} & \ldots & {\displaystyle
\frac{r_{2N}}{\a_{2N}-\beta_1}}\\ \vdots & \vdots & \ddots &
\vdots\\ \pm 1 & {\displaystyle \frac{r_1}{\a_1-\beta_N}} & \ldots
& {\displaystyle \frac{r_{2N}}{\a_{2N}-\beta_N}} \vspace{0.25cm}\\
0 & {\displaystyle \frac{h_1(\a_1)}{\a_1-\beta_1^*}} & \ldots &
{\displaystyle \frac{h_1(\a_{2N})}{\a_{2N}-\beta_1^*}}\\ \vdots &
\vdots & \ddots & \vdots\\ 0 & {\displaystyle
\frac{h_N(\a_1)}{\a_1-\beta_N^*}} & \ldots & {\displaystyle
\frac{h_N(\a_{2N})}{\a_{2N}-\beta_N^*}}\\
\end{array}\right|, \quad F=\left|\begin{array}{cccc} 0 & f(\a_1) & \ldots & f(\a_{2N})
\\ -1 & {\displaystyle \frac{r_1}{\a_1-\beta_1}} & \ldots &
{\displaystyle \frac{r_{2N}}{\a_{2N}-\beta_1}}\\ \vdots & \vdots &
\ddots & \vdots\\ -1 & {\displaystyle \frac{r_1}{\a_1-\beta_N}} &
\ldots & {\displaystyle \frac{r_{2N}}{\a_{2N}-\beta_N}}
\vspace{0.25cm}\\ 0 & {\displaystyle
\frac{h_1(\a_1)}{\a_1-\beta_1^*}} & \ldots & {\displaystyle
\frac{h_1(\a_{2N})}{\a_{2N}-\beta_1^*}}\\ \vdots & \vdots & \ddots &
\vdots\\ 0 & {\displaystyle \frac{h_N(\a_1)}{\a_1-\beta_N^*}} &
\ldots & {\displaystyle
\frac{h_N(\a_{2N})}{\a_{2N}-\beta_N^*}}\\
\end{array}\right|, \label{EF_2Ne} \eea
where the coordinates $\rho$ and $z$ enter the $(2N+1)\times(2N+1)$
determinants $E_\pm$ and $F$ only through the functions
$r_n=\sqrt{\rho^2+(z-\a_n)^2}$, and the constant objects $h_l(\a_n)$
and $f(\a_n)$ are defined as follows:
\bea\label{hfe} &&h_l(\a_n)=e_l^*+2f_l^* f(\a_n), \quad
f(\a_n)=\sum\limits_{l=1}^N\frac{f_l}{\a_n-\beta_l}, \nonumber\\
&&e_l=\frac{2\prod_{n=1}^{2N}(\beta_l-\a_n)} {\prod_{k\ne
l}^{N}(\beta_l-\beta_k)\prod_{k=1}^{N}(\beta_l-\beta_k^*)}
-2\sum\limits_{k=1}^N\frac{f_l f_k^*}{\beta_l-\beta_k^*}, \eea
the asterisk meaning complex conjugation. The set of arbitrary
parameters involved in formulas (\ref{EF}) and (\ref{hfe}) consists
of $N$ complex constants $\beta_l$, $N$ complex constants $f_l$ and
$2N$ real parameters supplied by the $\a_n$'s which can be
real-valued or occur in complex conjugate pairs.

Potentials $\E$ and $\Phi$ determine the functions $f$, $\gamma$ and
$\omega$ in the stationary axisymmetric line element
\be d s^2=f^{-1}[e^{2\gamma}(d\rho^2+d z^2)+\rho^2 d\varphi^2]-f(d
t-\omega d\varphi)^2, \label{Papa} \ee
and these are given by the formulas
\bea\label{metricN} f&=&\frac{D}{2E_-E_-^*}, \quad
e^{2\gamma}=\frac{D}{2K_0
K_0^*\prod_{n=1}^{2N}r_n}, \quad \omega=-\frac{2{\rm Im}[E_-^*(G+H)+FI^*]}{D}, \nonumber\\
D&=&E_+E_-^*+E_+^*E_-+2FF^*. \eea
The explicit form of the determinants $K_0$, $G$, $H$ and $I$ the
reader may find in Ref.~\cite{RMM}.

The subfamily describing $N$ arbitrary KN black holes located on the
symmetry axis and separated from each other by massless struts is
contained in the general formulas as a special asymptotically flat
case characterized by $4N-1$ real parameters representing the
individual $N$ masses, $N$ angular momenta, $N$ electric charges and
$N-1$ relative distances between the black holes. Moreover, since we
restrict ourselves to the black-hole sector of the solution
(\ref{EF}) only, then all $2N$ parameters $\a_n$ entering the
functions $r_n$ and determining the positions of black holes on the
symmetry axis must be real valued (see Fig.~1). At the same time,
the fact that the general $N$-soliton solution has $6N$ arbitrary
real parameters means that its multi-black hole subfamily we are
interested in arises by imposing $2N+1$ restrictions/conditions on
the parameters of the general solution. Such restrictions are quite
simple: the first one is just the condition of asymptotic flatness,
or absence of the NUT parameter of the system (one restriction);
furthermore, the separation of black holes means that the metric
function $\omega$ must verify the so-called {\it axis condition} on
the parts of the symmetry axis separating the black hole horizons
($N-1$ restrictions); of course, since we consider the conventional
KN black holes endowed with electric charges, we must also exclude
the individual magnetic charges of all the black holes ($N$ more
restrictions). To these $2N$ restrictions it is necessary to add the
last one related to the possibility of a translation along the
$z$-axis, and this liberty will be abolished by fixing the position
of the origin of coordinates. In mathematical terms, the above
restrictions can be formulated as
\be\label{af} {\rm Im}\left(\sum\limits_{l=1}^N e_l\right)=0 \ee
(the condition of asymptotic flatness),
\be\label{axis} \omega(\rho=0,\a_{2k+1}\le z\le\a_{2k})=0, \quad
k=1,2,...,N-1 \ee
(the axis conditions), where one must bear in mind that $\omega$
takes constant values on the symmetry axis, and
\be\label{zmc} {\rm Re}[\Phi(\rho=0,z=\a_{2k-1})-
\Phi(\rho=0,z=\a_{2k})]=0, \quad k=1,2,...,N \ee
(absence of magnetic charges). Lastly, the position of the origin of
coordinates can be fixed, say, by subjecting $\a_n$'s to the
constraint
\be\label{alf} \sum\limits_{n=1}^{2N}\a_n=0, \ee
which turns out to be particularly useful in the case of the
equatorially symmetric configurations.

To calculate the Euclidean action for the system of $N$ collinear KN
black holes, we must perform a Wick rotation in the metric
(\ref{Papa}), yielding
\begin{equation}\label{PapaW}
    d\tilde s^2=f^{-1}\left[e^{2\gamma}(d\rho^2+dz^2)
    +\rho^2 d\varphi^2\right]-f(id\tau+\omega d\varphi)^2,
\end{equation}
and then follow the calculation procedures outlined in the previous
section. We shall compute the terms $I_{\rm EH}$, $I_{\rm GH}$ and
$I_{\rm em}$ in the same order as this was done in the case of a
single KN black hole.

\subsection{The Einstein-Hilbert term}

In the general case of $N$ black holes, two groups of conical
singularities are present: the first group is comprised of $N$
singularities on the horizons of black holes as a consequence of the
Wick rotation, and the second group accounts for $N-1$ conical
singularities due to the usual massless struts that prevent the
black holes from falling onto each other. Although these types of
singularities look different, their contribution to the action
integral $I_{\rm EH}$ must be evaluated by means of the same formula
(\ref{cs}). Below we will start with the conical singularities
arising from the struts.

To compute the $Area$ $A^S_i$ associated to ${\cal L}_{i}\times
S^1$, where ${\cal L}_i$ is the $i$-th strut joining the points
$\a_{2i}$ and $\a_{2i+1}$ of the symmetry axis and $S^1$ refers to
the periodic time, we can use the formula
\be\label{al}
    A^S_i=\int_{{\cal L}_i\times S^1}\sqrt{h},
\ee
where $h$ is the determinant of the induced metric over the
hypersurface ${\cal L}_{i}\times S^1$. From (\ref{PapaW}) it follows
that $h=\exp(2\gamma)$, and we obtain
\be\label{ali}
    A^S_i=\int_{\a_{2i+1}}^{\a_{2i}}
    \int_0^\beta e^\gamma |_{{\cal L}_{i}}dzd\tau
    =\beta e^{\gamma_i}L_i,
\ee
where $L_i=\a_{2i}-\a_{2i+1}$ is the coordinate length of the $i$-th
strut and $\gamma_i$ is the (constant) value of the metric function
$\gamma$ on the $i$-th strut. Introducing further the {\it
thermodynamic length} \cite{AGK} of the $i$-th strut by the formula
$l_i=\exp(\gamma_i)L_i$, \cite{HRR,KZe} we finally get
\be\label{abl}
    A^S_i=\beta l_i.
\ee

On the other hand, the corresponding deficit angle related to the
coordinate $\varphi$ can be written as
\be\label{df}
    \delta^\varphi_i=2\pi-\int^{2\pi}_0
    e^{-\gamma_i}d\varphi=2\pi(1-e^{-\gamma_i})=-8\pi\mathcal{F}_{i},
\ee
where $\mathcal{F}_{i}$ represents the interaction force between the
$i$-th and $(i+1)$-th black holes \cite{Isr,Wei2}
\be\label{Fi} \mathcal{F}_{i}=\frac{1}{4}(e^{-\gamma_i}-1), \quad
i=1,2,...,N-1. \ee

As for the $N$ conical singularities generated by the Euclidean time
$\tau$ on the black hole horizons, it was already mentioned in the
previous section that their $Area$'s entering formula (\ref{cs}) are
just areas of black holes $A^H_i$, and technically these can be
computed by means of the formula
\be\label{ahi}
    A^H_i=\int_{H_i}\sqrt{h},
\ee
where $h=-e^{2\gamma}\omega^2$, as it follows from (\ref{PapaW});
then
\be\label{ahis}
    A^H_i=\int_{\a_{2i}}^{\a_{2i-1}}
    \int_0^{2\pi} \sqrt{-e^{2\gamma_i}\omega_i^2}\ dz\,d\varphi
    =4\pi\sigma_i\sqrt{-e^{2\gamma_i}\omega_i^2}.
\ee
Here $\gamma_i$ and $\omega_i$ are the constant values of the metric
functions $\gamma$ and $\omega$ on the $i$-th horizon, and
$\s_i=(\a_{2i-1}-\a_{2i})/2$ is the half length of the $i$-th
horizon. Formula (\ref{ahis}) can also be rewritten in the form
\be\label{ahik} A^H_i=4\pi\sigma_i\kappa_i^{-1}, \ee
where $\kappa_i=(-e^{2\gamma_i}\omega_i^2)^{-1/2}$ is the surface
gravity of the $i$-th black hole horizon \cite{Car,Tom}.

For the corresponding deficit angle $\delta^\tau_i$ related to the
$i$-th horizon and associated with the periodic time $\tau$ we have
\be\label{dti}
    \delta^\tau_i=2\pi-\int^\beta_0 \kappa_id\tau
    =2\pi\left(1-\frac{\kappa_i}{2\pi}\beta\right),
\ee
and therefore the expression of the action term $I_{\rm EH}$ takes
the form
\bea\label{iehN}
    I_{\rm EH}&=&-\frac{1}{8\pi}\sum^{N-1}_{i=1}
    A^S_i\delta^\varphi_i
    -\frac{1}{8\pi}\sum^N_{i=1}A^H_i\delta^\tau_i \nonumber\\
    &=&\beta\sum^{N-1}_{i=1} l_i\mathcal{F}_{i}
    -\frac{1}{4}\sum^N_{i=1}A^H_i
    \left(1-\frac{\kappa_i}{2\pi}\beta\right).
\eea

\subsection{The Gibbons-Hawking term}

In order to calculate the Gibbons-Hawking term (\ref{igh}) of the
Euclidean action for our multi-black hole system, it is convenient
to make use of spherical coordinates $(\zeta,\theta)$ related to the
Weyl-Papapetrou cylindrical coordinates $(\rho,z)$ by the formulas
\be
    \zeta=\sqrt{\rho^2+z^2}, \quad
    \cos\theta=z/\sqrt{\rho^2+z^2}.
\ee
Then, by analogy with the case of a single KN black hole, we can
consider a hypersurface $\mathcal{N}$ defined by $\zeta={\rm
const}$. Clearly, as $\zeta\to\infty$, the asymptotic behavior of
the functions $r_n$ of the solution (\ref{EF}) is $r_n\to\zeta$ .

Since the spacetime under consideration is asymptotically flat, then
the component $g_{\tau\varphi}$ of the metric tensor is of the order
$O(\zeta^{-1})$, and the induced metric over $\mathcal{N}$ can be
written in the form
\be\label{mi}
    d\sigma^2=f_{\mathcal{N}}d\tau^2+f^{-1}_{\mathcal{N}}\zeta^2
    \left(e^{2\gamma_\mathcal{N}}d\theta^2+\sin^2{\theta}
    d\varphi^2\right)
    +O\left(\zeta^{-1}\right)d\tau d\varphi,
\ee
with the determinant
\be\label{det}
    h=
    \left|\begin{array}{cccc}
    f_{\mathcal{N}}&0&0\\
    0&f^{-1}_{\mathcal{N}}\zeta^2\sin^2{\theta}&O\left(\zeta^{-1}\right)\\
    0&O\left(\zeta^{-1}\right)&f^{-1}_{\mathcal{N}}\zeta^2 e^{2\gamma_\mathcal{N}}
    \end{array}\right|
    =f^{-1}_{\mathcal{N}}e^{2\gamma_\mathcal{N}}\zeta^4\sin^2{\theta}
    +O\left(\zeta^{-2}\right),
\ee
where the subscript $\mathcal{N}$ indicates that the functions are
restricted to $\mathcal{N}(\zeta)$.

We note that $I_{\rm GH}$ is defined at spatial infinity, so we are
free to use a large-$\zeta$ approximation for our purposes, within
which we can neglect the terms of order $O(\zeta^{-2})$. In
particular, we will drop for that reason the last term in
(\ref{det}) involving imaginary quantities. Computing now the volume
of $\mathcal{N}$ for large $\zeta$, we obtain
\be\label{VN}
    {\rm Vol}\,\mathcal{N}=\int_\mathcal{N}\sqrt{h}\, d\tau d\varphi d\theta
   \simeq 4\pi\beta\zeta^2f_\mathcal{N}^{-1/2}e^{\gamma_\mathcal{N}},
\ee
where it has been supposed that $f_\mathcal{N}$ as well as
$\gamma_\mathcal{N}$ are constant on $\mathcal{N}(\zeta)$.

The asymptotic behavior of the function $f_\mathcal{N}$ of the
solution (\ref{EF}) is well known, and it is determined by the
expression
\be\label{fNa}
    f_\mathcal{N}\simeq 1-\frac{2M}{\zeta}, \quad M=-\frac{1}{2}{\rm
    Re}\left(\sum\limits_{l=1}^Ne_l\right),
\ee
where $M$ is the total mass of the system. In the particular case of
the black hole spacetime, the struts give zero contribution to the
total mass, the latter then representing a sum of individual Komar
masses \cite{Kom} of the black hole constituents
\be\label{mtK} M=\sum\limits_{i=1}^N M_i. \ee
Of course, $M$ here can also be viewed as the ADM mass of the
multi-black hole system \cite{ADM}.

Now, taking into account (\ref{fNa}), we get
\bea\label{gVN}
    g_{\zeta\zeta}^{-1/2}\partial_\zeta {\rm Vol}\,\mathcal{N}&\simeq&4\pi\beta
    \left(1+\frac{2M}{\zeta}\right)^{-1/2}\partial_\zeta
    \left[\left(1+\frac{2M}{\zeta}\right)^{1/2}\zeta^2\right] \nonumber\\
    &=&4\pi\beta\left(2\zeta-M\right)+O\left(\zeta^{-1}\right),
\eea
whence it follows that
\be\label{iNK}
    \int_\mathcal{N} K\sqrt{h}=g_{\zeta\zeta}^{-1/2}\partial_\zeta
    {\rm Vol}\,\mathcal{N}=4\pi\beta
    \left(2\zeta-M\right)+O\left(\zeta^{-1}\right).
\ee

In addition,
\be\label{iNK0}
    \int_\mathcal{N} K_0\sqrt{h}=4\pi\beta(2\zeta)
    +O\left(\zeta^{-1}\right),
\ee
and hence
\be\label{iKd}
    \int_\mathcal{N} \left[K\right]\sqrt{h}=-4\pi\beta M
    +O\left(\zeta^{-1}\right).
\ee

Therefore, after taking the limit $\zeta\to\infty$ in (\ref{iKd}),
we finally arrive at
\bea\label{iGHN}
    I_{\rm GH}=\frac{1}{2}\beta M.
\eea

Of course, this result is an expected one, and in the case of
asymptotically flat spacetimes it may even look trivial because all
it says is that the total mass of the system is a combined
contribution of all the constituents of the system. However, the
calculation of the total mass can become a nontrivial exercise if a
spacetime is not asymptotically flat globally \cite{GLS}.

\subsection{The electromagnetic term}

Like in the case of a single KN black hole, the electromagnetic
potential $A$ of the generic configuration of $N$ black hole
constituents has two nonzero components and is defined by the
formula
\be\label{AN}
    A=A_\tau d\tau+A_\varphi d\varphi,
\ee where the explicit form of $A_\tau$ and $A_\varphi$ is given in
the paper \cite{RMM}.

To compute the electromagnetic term $I_{\rm em}$ of the Euclidean
action, it is advantageous first to rewrite (\ref{iem}) as a
boundary integral
\be\label{Ib}
    I_{\rm em}=\frac{1}{8\pi}\int_{\partial\mathcal{M}}A\wedge\star
    dA,
\ee
also noting that since the field $A$ vanishes at infinity we have
$\partial M=\sum_i H_i\times S^1$, $H_i$ denoting as usual the
horizon of $i$-th KN black hole.

To perform the calculation, we introduce a family of $N$
hypersurfaces $N_i=C_i\times S^1$, where $C_i$ is a cylinder with
radius $\rho={\rm const}$ and height $\a_{2i-1}-\a_{2i}$. In this
way we will have the behavior $N_i\to H_i\times S^1$ as $\rho\to 0$.
Since the integrals over the bases of the cylinders do not
contribute in the final result, the computation for each $N_i$
yields
\be\label{emN}
    A\wedge\star dA|_{N_i}=\rho^{-1}\left[A_\tau(g_{\varphi\tau}
    \partial_\rho A_\varphi-g_{\varphi\varphi}
    \partial_\rho A_\tau)
    +A_\varphi(g_{\tau\varphi}
    \partial_\rho A_\tau-g_{\tau\tau}\partial_\rho A_\varphi)
    \right]d\tau\wedge dz\wedge
    d\varphi|_{N_i}.
\ee

Following Carter \cite{Car}, we now define
$\lambda_i:=-i\partial_\tau+\Omega^H_{i}\partial_{\varphi}$ and
$\Omega^H_{i}:=-i g_{\tau\varphi}/g_{\varphi\varphi}|_{H_i}$. Then
we can rewrite the first $A_\tau$ in (\ref{emN}) as
$A_\tau-i\Omega^H_{i}A_\varphi+i\Omega^H_{i}A_\varphi$ and rearrange
the terms, thus obtaining
\bea\label{ANi}
    \int_{N_i}A\wedge\star dA&=&\beta\int_{N_i-S^1}\rho^{-1}
    (A_\tau-i\Omega^H_{i}A_\varphi)\left(g_{\varphi\tau}
    \partial_{\rho}A_\varphi-g_{\varphi\varphi}\partial_{\rho}
    A_\tau\right)dz d\varphi \nonumber\\
    &&+2\pi\beta\int^{\a_{2i-1}}_{\a_{2i}}\rho^{-1} A_\varphi \left[(g_{\tau\varphi}-i\Omega^H_{i}
    g_{\varphi\varphi})\partial_\rho A_\tau-(g_{\tau\tau}-i\Omega^H_{i}g_{\varphi\tau})
    \partial_\rho A_\varphi\right]dz. \nonumber\\
\eea

Observing further that
\be\label{appr}
    g_{\tau\varphi}-i\Omega^H_{i}g_{\varphi\varphi}\approx 0, \quad
    g_{\tau\tau}-i\Omega^H_{i}g_{\varphi\tau}\approx\rho^2
    g_{\varphi\varphi}^{-1}
\ee
in the vicinity of the $i$-th horizon, and also that $\rho\to0$ as
$N_i\to H_i\times S^1$, with which the second integral on the
right-hand side of (\ref{ANi}) vanishes on the horizon, we get
\bea\label{ANiL}
    \int_{H_i\times S^1}A\wedge\star dA&=&
    \lim_{\rho\to 0}\int_{N_i}A\wedge\star dA \nonumber\\
    &=&\beta i\Phi_i\int_{H_i}\star F +2\pi\beta\lim_{\rho\to 0}
    \int^{\a_{2i-1}}_{\a_{2i}} \rho^{-1}A_\varphi
    [(g_{\tau\varphi}-i\Omega^H_{i}g_{\varphi\varphi})\partial_\rho
    A_\tau \nonumber\\ &&-(g_{\tau\tau}-i\Omega^H_{i}g_{\varphi\tau})\partial_\rho
    A_\varphi]dz, \nonumber\\ &=&\beta \Phi_i \left(i\int_{H_i}\star
    F\right),
\eea
where we have exploited the fact that the potential
\be\label{Fi}
\Phi_i:=A(\lambda_i)|_{H_i}=iA_\tau+\Omega^H_{i}A_\varphi|_{H_i} \ee
takes constant value on the $i$-th horizon \cite{Car}, and also made
use of the relation
\be\label{relFA}
    \star F|_{H_i}=\star dA|_{H_i}
    =-\rho^{-1}\left(g_{\varphi\tau}\partial_{\rho}A_\varphi
    -g_{\varphi\varphi}\partial_{\rho}
    A_\tau\right)dz\wedge d\varphi|_{H_i}.
\ee

Last, using formula \cite{Car}
\be\label{gau}
    i\int_{H_i}\star F=-4\pi Q_i,
\ee
where $Q_i$ is the electric charge of the $i$-th KN black hole, we
arrive at the desired result for the electromagnetic action term:
\be\label{Rsem}
    I_{\rm em}=-\frac{1}{2}\beta\sum^{N}_{i=1} Q_i\Phi_i.
\ee

Putting now the expressions obtained for $I_{\rm EH}$, $I_{\rm GH}$
and $I_{\rm em}$ together, we can write down the final formula for
the Euclidean action $I_{\rm E}$ of $N$ arbitrary KN black holes:
\be\label{IeN} I_{\rm E}=\frac{\beta}{2}\left(M-\sum^{N}_{i=1}
Q_i\Phi_i\right) +\beta\sum^{N-1}_{i=1} l_i\mathcal{F}_{i}
    -\frac{1}{4}\sum^N_{i=1}A^H_i
    \left(1-\frac{\kappa_i}{2\pi}\beta\right). \ee

It is clear that the last term in (\ref{IeN}) involving $A^H_i$ and
$\kappa_i$ is likely to be removed from the action for the
regularity reason.

\section{Hawking average temperature and free energy of $N$ KN black holes}

We have seen that in the case of a single KN black hole the conical
singularity formed on the horizon by the periodic time can be
eliminated by an appropriate choice of $\beta$, which is equivalent
to introducing the Hawking temperature of a black hole. However, in
the configurations consisting of various black holes which have
different surface gravities, the conical singularity can be removed
(by fixing $\beta$) only on one horizon, as it follows from the
structure of the last term in (\ref{IeN}), whereas the horizons of
the other black holes will remain singular. This in particular
explains why the Euclidean action method has been applied so far
exclusively to the binary systems of equal black holes.

At the same time, the advantage of evaluating explicitly the
singular terms in the Euclidean action is quite clear -- this opens
a good opportunity to analyze these terms in detail, and helps one
understand better the whole problem with the conical singularities
due to Euclidean time and possible approaches to tackle it. Thus, an
immediate proposal for regularizing the Euclidean action would be
demanding that the entire combined contribution of the singular
horizon terms in (\ref{IeN}) must vanish. As can be easily seen,
this fixes {\it uniquely} the choice of the parameter $\beta$ in the
action; indeed,
\be\label{bf} \sum^N_{i=1}A^H_i
    \left(1-\frac{\kappa_i}{2\pi}\beta\right)=0 \quad
    \Longrightarrow \quad \beta\equiv\hat\beta=\left(\sum^N_{i=1} A^H_i\right)
    \left(\sum^N_{i=1}\frac{\kappa_i}{2\pi}A^H_i\right)^{-1}, \ee
whence we get the expression for the temperature that generalizes
the notion of the Hawking temperature to the case of multiple black
holes:
\be\label{hat} T\equiv\hat T=\frac{1}{2\pi}\left(\sum^N_{i=1}
\kappa_i A^H_i\right)\left(\sum^N_{i=1} A^H_i\right)^{-1}. \ee
From now on we will call $\hat T$ the {\it Hawking average
temperature} (HAT) of a multi-black hole system.

Note that in the particular cases of one black hole
($\kappa_1=\kappa$, $\kappa_i=0$, $i=2,...,N$), or a system of black
holes possessing the same surface gravity ($\kappa_i=\kappa$,
$i=1,...,N$), one recovers from (\ref{hat}) the well-known
conventional formula for the Hawking temperature:
\be\label{TH}
    T_{\rm H}=\frac{\kappa}{2\pi}. \ee

Therefore, for the regularized Euclidean action in the presence of
massless struts we obtain the formula
\be\label{IeR} I_{\rm E}=\frac{\hat\beta}{2}\left(M-\sum^{N}_{i=1}
Q_i\Phi_i\right) +\hat\beta\sum^{N-1}_{i=1} l_i\mathcal{F}_{i}, \ee
which in turn supplies us, via (\ref{fe}), with the expression for
the {\it free energy} of the system:
\be\label{feN}
     W=\frac{1}{2}\left(M-\sum^N_{i=1}
     Q_i\Phi_i\right)+\sum^{N-1}_{i=1} l_i\mathcal{F}_i.
\ee

We would like to emphasize that the introduction of HAT $\hat T$
does not actually deny anyhow the existence of individual Hawking
temperatures $T_i$ of the black hole constituents. Precisely the
possibility of removing the conical singularity on any particular
horizon by means of the choice $\beta_i=1/T_i=2\pi/\kappa_i$ shows,
on the one hand, that the individual black hole temperatures do
exist and are of Hawking's type and, on the other hand, that they
may not coincide with the average temperature $\hat T$ of the
system. In this respect it should be pointed out that we assume as
usual that each black hole satisfies the Smarr mass relation
\cite{Sma}, and the sum of such relations can be written as
\be\label{Sma}
M=\sum^N_{i=1}M_i=\sum^N_{i=1}\left(2T_iS_i+2\Omega^H_iJ_i+\Phi_iQ_i
\right), \ee
where $J_i$ is the angular momentum and $S_i$ the entropy of the
$i$-th black hole, while $\Omega^H_i$ introduced in the previous
section represents the angular velocity of the $i$-th horizon and is
equal to the inverse of the metric function $\omega$ evaluated on
that horizon. In fact, it is not difficult to find a representation
of $\hat T$ in terms of the individual Hawking temperatures $T_i$.
Indeed, performing in (\ref{hat}) the substitutions
$\kappa_i=T_i/2\pi$, $A^H_i=4S_i$, and introducing the total entropy
$S$ of the system by the formula
\be\label{St}
    S:=\sum^N_{i=1}S_i=\sum^N_{i=1}\frac{A^H_i}{4},
\ee
we get for $\hat T$ the following concise expression:
\be\label{TTi} \hat T=\frac{1}{S}\sum^N_{i=1}T_i S_i, \ee
with which for instance the Smarr relation (\ref{Sma}) rewrites as
\be\label{SmaT} M=2\hat T
S+\sum^N_{i=1}\left(2\Omega^H_iJ_i+\Phi_iQ_i \right). \ee

A useful corollary of (\ref{SmaT}) is yet another form of the free
energy, namely,
\be\label{feT}
    W=M-\hat T S
    -\sum^N_{i=1}\left(\Omega^H_i J_i+\Phi_i Q_i\right)
    +\sum^{N-1}_{i=1}l_i\mathcal{F}_i,
\ee
which generalizes the respective formula for $W$ known for the case
of two equal black hole constituents \cite{CPe,GMR}.

As was shown in \cite{GMR}, the variation of $W$ is given by the
formula
\be\label{gmr}
    dW=-Sd\hat T-\sum^N_{i=1}(J_id\Omega^H_i+Q_id\Phi_i)
    +\sum^{N-1}_{i=1}\mathcal{F}_idl_i,
\ee
so by taking the differential of (\ref{feT}) and making use of
(\ref{gmr}) we obtain the first law of thermodynamics for the system
of $N$ KN black holes:
\be\label{flt}
    dM=\hat TdS+\sum^N_{i=1}(\Omega^H_idJ_i+\Phi_idQ_i)
    -\sum^{N-1}_{i=1}l_id\mathcal{F}_i.
\ee

Returning now to the temperature $\hat T$, let us consider its
representation involving the quantities $\s_i$ that are defined by
the parameters $\a_n$ of the $N$-soliton solution and represent the
half lengths of the respective horizons. As it follows from the
formula (\ref{ahik}), we have $\kappa_iA^H_i=4\pi\s_i$, so that,
after changing the sum of the horizon areas to the sum of the
entropies, we can rewrite (\ref{hat}) in the form
\be\label{Ths} \hat T=\frac{1}{2S}\sum^N_{i=1}\s_i. \ee

The above representation of $\hat T$ is of interest for two reasons.
First, it demonstrates in a simple way that in the limiting case of
$N$ extreme KN black holes (when all $\s_i=0$) the HAT $\hat T$
takes zero value. Second, formula (\ref{Ths}) simplifies even
further for a static configuration of $N$ Schwarzschild black holes
\cite{IKh}, the thermodynamics of which has recently been studied in
the papers \cite{GLS,LPo1,LPo2}, because in this case the quantities
$\s_i$ are equal to the masses of black holes, $\s_i=M_i$, and, as a
consequence, (\ref{Ths}) takes a remarkably simple form
\be\label{ThM} \hat T=\frac{M}{2S}, \ee
where $M$ is the total mass of the Schwarzschild black holes.

In the simplest case of two nonequal Schwarzschild black holes
\cite{BWe}, the explicit form of $\hat T$ can be worked out with the
aid of the formulas of paper \cite{MRS}, yielding
\be\label{T2S} \hat T=\frac{M[R^2-(M_1-M_2)^2]}
{8\pi(R+M)[R(M_1^2+M_2^2)-M(M_1-M_2)^2]}, \quad M=M_1+M_2, \ee
where $R$ is the coordinate distance between the centers of black
holes. If $M_1>M_2$, then $\hat T$ turns out to be greater than the
Hawking temperature of the black hole with larger mass $M_1$ and
less than the temperature of the black hole with smaller mass $M_2$:
\be\label{Tin} \frac{R+M_1-M_2}{8\pi M_1(R+M)} < \hat T <
\frac{R-M_1+M_2}{8\pi M_2(R+M)}, \ee
which lends support to our interpretation of $\hat T$ as an average
temperature of the system. Note that $\hat T$ is an increasing
function of $R$, taking its maximum value,
\be\label{Tl} \hat T_{\rm max}=\frac{M}{8\pi(M_1^2+M_2^2)}, \ee
in the limit $R\to\infty$, and its minimum value $T_{\rm min}=(8\pi
M)^{-1}$ at $R=M$, when the two black holes merge and form one
Schwarzschild black hole with mass $M$. The gravitational
interaction of black holes therefore decreases the temperature of
the system!

\section{Concluding remarks}

Therefore, we have computed the Euclidean action describing the
system of $N$ arbitrary KN black holes and shown how the problem of
singular horizon terms due to periodic time can be obviated by
introducing the notion of average temperature of a multi-black hole
configuration. In this way we have extended the applicability of the
path-integral method to the systems of nonequal black holes with
different surface gravities, which in particular permitted us to
obtain the expression for the free energy of the system in the
generic case and find a natural generalization of the Hawking
temperature to the case of multiple black holes in the form of HAT
$\hat T$ determined concisely by the surface gravities and horizon
areas of black holes. It is worth noting in this respect that the
Euclidean action method seems to be ideally suited for treating the
black hole systems as it permits a far-reaching analysis just on the
basis of the well-known general properties and characteristics of
black holes, without the need to use the explicit form of the
multi-black hole solution.

While the average temperature $\hat T$ introduced in the present
paper offers a possibility to approximate a system of black holes by
treating it as one body with mass $M$ (the total mass of the system)
and temperature $\hat T$, one may also ask a question whether the
system could have some other averaged physical characteristics
describing it as an integral unit. Interestingly, looking at the
Smarr relation in the form (\ref{SmaT}), it is tempting to speculate
that the second and third terms on the right-hand side of
(\ref{SmaT}) could be rewritten similar to the first term involving
the total entropy. Indeed, by introducing the total angular momentum
$J$ and total charge $Q$ of the system by the usual formulas
\be\label{JQ} J=\sum^N_{i=1}J_i, \quad Q=\sum^N_{i=1}Q_i, \ee
we can define, in analogy with the HAT $\hat T$, the average angular
velocity $\hat\Omega$ and average electric potential $\hat\Phi$ as
\be\label{JQ} \hat\Omega=\frac{1}{J}\sum^N_{i=1}\Omega^H_iJ_i, \quad
\hat\Phi=\frac{1}{Q}\sum^N_{i=1}\Phi_i Q_i, \ee
with which the Smarr relation (\ref{SmaT}) takes the form
\be\label{SmaH} M=2\hat TS+2\hat\Omega J+\hat\Phi Q. \ee
The only reason why this formula has not received attention in the
literature so far is of course the necessity to justify $\hat T$ as
the average temperature of the system, something that slightly
surpasses the limits of the classical black hole thermodynamics. We
hope that our paper providing such a justification for $\hat T$ by
means of the Euclidean action approach makes formula (\ref{SmaH})
fully plausible now.

\section*{Acknowledgments}

This work was partially supported by Project~128761 from CONACyT of
Mexico. C.J.R.V. acknowledges financial support from CONACyt of
Mexico, Grant No.~278847.

\newpage

\begin{figure}[htb]
\centerline{\epsfysize=75mm\epsffile{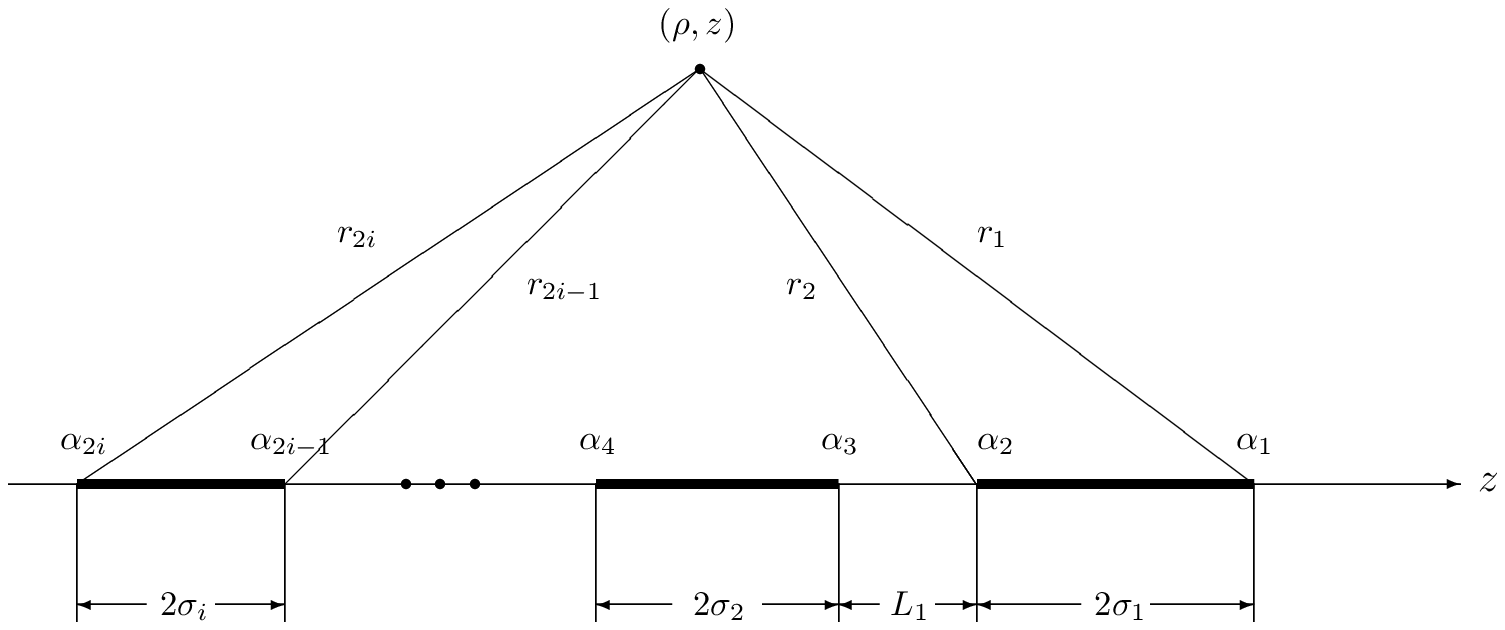}} \caption{Location of
the KN black holes on the symmetry axis.}
\end{figure}

\end{document}